\documentclass[aps,prd,twocolumn,superscriptaddress,preprintnumbers,floatfix,nofootinbib,notitlepage,showkeys,showpacs]{revtex4-1}

\usepackage[utf8x]{inputenc}
\usepackage{graphicx}
\usepackage{hyperref}
\usepackage{latexsym}
\usepackage{amsmath}
\usepackage{amssymb}
\usepackage{bbm}

\usepackage{ulem}
\usepackage{pdfsync}
\usepackage{epsfig}
\usepackage{epstopdf}

\newcommand{\A}{a}

\newcommand{\ev}[1]{\left \langle #1 \right \rangle}

\newcommand{\csw}{c_{\rm{sw}}}

\newcommand{\Dslash}{\ensuremath \raisebox{0.025cm}{\slash}\hspace{-0.27cm} D}
\newcommand{\be}{\begin{equation}}
\newcommand{\ee}{\end{equation}}
\newcommand{\bea}{\begin{eqnarray}} 
\newcommand{\eea}{\end{eqnarray}}

\newcommand{\bmp}{\noindent\begin{minipage}{16cm}}
\newcommand{\emp}{\end{minipage}\vskip 7mm} 
\def\lsim{\mathrel{\raise.3ex\hbox{$<$\kern-.75em\lower1ex\hbox{$\sim$}}}}
\def\gsim{\mathrel{\raise.3ex\hbox{$>$\kern-.75em\lower1ex\hbox{$\sim$}}}}

\newcommand{\intron}[1]{}

\begin{document}

\title{ Gradient Flow Coupling in the SU(2) gauge theory with two adjoint fermions}
  

\author{Jarno Rantaharju}
\email{rantaharju@cp3.sdu.dk}
\affiliation{CP-Origins \& IMADA, Campusvej 55, DK-5230 Odense M, Denmark\\ and RIKEN Advanced Institute for Computational Science, Kobe, Hyogo 650-0047, Japan}

\begin{abstract}
We study SU(2) gauge theory with two fermion flavors in the adjoint representation. Using a clover improved HEX smeared action and the gradient flow running coupling allows us to simulate with larger lattice size than before. We find an infrared fixed point after a continuum extrapolation in the range $4.3 \lsim g^{*2} \lsim 4.8 $. We also measure the mass anomalous dimension and find the value $ 0.25 \lsim \gamma^* \lsim 0.28 $ at the fixed point.
\end{abstract}

\preprint{CP3-Origins-2015-050 DNRF90 }


\maketitle

\section{Introduction}

The quantitative determination of the phase space of SU($N$) gauge theories coupled to $N_f$ fermions in different representations of the gauge field provides a challenge in nonperturbative physics. The loss of asymptotic freedom when the number of fermion flavors is large can be understood perturbatively. Asymptotic freedom simply depends on the sign of the lowest order term in the perturbative expansion of the $\beta$-function \cite{Banks:1981nn}. Below the loss of asymptotic freedom there is a range of $N_f$ where the theory has a non-trivial infrared fixed point (IRFP). With a sufficiently small number of fermion flavors the model develops a chiral condensate, which dominates its infrared behavior.
Whether the theory is chirally broken or has an IRFP depends on its behavior in the deep infrared where the coupling can be large. The question is therefore nonperturbative in nature.

The existence of gauge theories with significantly different dynamics from QCD has in the recent years generated significant interest. Infrared conformal models are also of interest for model building beyond the Standard Model. A good example is provided by technicolor theories, where the electroweak symmetry is broken by the formation of a chiral condensate in a strongly interacting sector \cite{TC,Eichten:1979ah,Hill:2002ap,Sannino:2008ha}. Various approximations can be used to study the lower limit of the conformal window \cite{Sannino:2004qp}, but lattice simulations provide the only first principles method of studying the nonperturbative dynamics of these theories.

In this work we study the SU(2) gauge theory coupled to 2 flavors of fermions transforming according to the adjoint representation of the gauge field. The model has been studied in previous works by several groups \cite{Catterall:2007yx,Hietanen:2008mr,Hietanen:2009az,DelDebbio:2008zf,Catterall:2008qk,
    Bursa:2009we,DelDebbio:2009fd,DelDebbio:2010hx,
    DelDebbio:2010hu,Bursa:2011ru,DeGrand:2011qd, Patella:2012da,
    Giedt:2012rj,Rantaharju:2013bva, Bergner:2015jdn,
    Rantaharju:2015yva,DelDebbio:2015byq}.
It is worth noting that results from this model can be used to constrain models with fermions in 2-index representations in a large $N_c$ approximation \cite{Bergner:2015dya}.
The model is likely to have an IRFP, but numerical estimates of scheme independent quantities such as the anomalous dimensions of the mass and the coupling carry large numerical and systematic uncertainties.

We employ the gradient flow method \cite{Luscher:2010iy,Luscher:2011bx,Fodor:2012td} and measure the running coupling using a larger lattice size than before.
We use the same action as in \cite{Rantaharju:2015yva},
where the coupling was measured in the Sch\"odinger functional scheme.
Since the two methods use different renormalization schemes, the values are not directly comparable.
Scheme independent quantities, however, can be compared.

We measure the mass anomalous dimension at the fixed point indicated by the running of the gradient flow coupling using the data first reported in \cite{Rantaharju:2015yva}. We find the value $\gamma^*\sim 0.27$. The result is larger than the one obtained in \cite{Rantaharju:2015yva}, indicating systematic uncertainty, possibly due to the continuum extrapolation of the coupling. In section \ref{coupling} of this paper we introduce the model and describe the measurement of the running coupling. We describe the measurement of the mass anomalous dimension in section \ref{gamma} and conclude in section \ref{conclusions}.

\section{The Running Coupling} \label{coupling}

We use a partially smeared Wilson plaquette action 
and the clover improved Wilson fermion action with
hypercubic truncated stout smearing (HEX smearing) \cite{Capitani:2006ni,Shamir:2010cq}.
The smearing helps to reduce the discretization errors and allows simulations at larger couplings
than an unsmeared action does \cite{DeGrand:2011qd}.
The action is the same as in \cite{Rantaharju:2015yva}.

The gauge action is a mixture of a single-plaquette Wilson action constructed out of smeared gauge field $V$ and unsmeared gauge field $U$:
\begin{align}
  S_G
  &= \beta_L \sum_{x;\mu<\nu} (1-c_g) {\mathcal L}_{x,\mu\nu}(U) + c_g {\mathcal L}_{x,\mu\nu}(V) 
   \label{plaqaction} 
\end{align}
where $\beta_L = 4/g_0^2$ and ${\mathcal L}_{x,\mu\nu}(U)$ is the Wilson gauge action for the field $U$. The properties of the gauge action are not sensitive to the precise value of $c_g$ and here we use $c_g=0.5$. 

The fermions belong to the adjoint representation of SU(2).  
We use the Wilson-clover fermion action
\begin{align}
  S_F
  & = a^4\sum_x \bigg [
  \bar{\psi}(x) ( i\, \Dslash_W + m_0 )
  \psi(x)  \nonumber \\
 & + a \csw \bar\psi(x)\frac{i}{4}\sigma_{\mu\nu}
  F_{\mu\nu}(x)\psi(x) \bigg ],
\end{align}
where $\Dslash_W$ is the standard Wilson Dirac operator.  The gauge link
matrices appearing in $S_F$ are in the adjoint
representation and are constructed from the smeared matrices $V_{x,\mu}$.
We use the tree-level clover coefficient $\csw=1$, which is expected to be a good approximation 
with smeared gauge links \cite{Capitani:2006ni,Shamir:2010cq,DeGrand:2011qd}.
The full action is then parametrized by the bare coupling $\beta_L=4/g_0^2$ and the hopping parameter $\kappa = 1/(2m_0 + 8)$.

The Wilson fermion action breaks chiral symmetry and requires additive renormalization of the quark mass. Thus, in order to simulate the massless theory, we need to determine the critical bare mass where the physical quark mass vanishes.  
We define the quark mass $M$ 
through the lattice partially conserved axial current PCAC relation
\begin{align}
aM(t) &= \frac14\frac{f_A(t+a)-f_A(t-a)}{f_P(t)}
\end{align}
and we define $\kappa_c$ as the value of the parameter $\kappa$ where $M(t=L/2)$ vanishes.
The values of $\kappa_c$ used in the simulations are given in table \ref{table:kappac}.

\begin{table}
\centering
\begin{tabular}{llll}
\hline
 $\beta_L$ & $\kappa_c$ & $aM(L/2)$ & $N_{traj}$    \\ \hline
\hline
 8    & 0.125842  & -6.3(3)e-5  & 14647 \\   
 4    & 0.127352  & 1(8)e-6  & 12961 \\ 
 2    & 0.132309  & -8(3)e-5 & 11797 \\ 
 1.5  & 0.136362 & -7.3(3)e-4    & 11026 \\ 
 1.3  & 0.13903   & -1.38(6)e-3  & 10797 \\ 
 1.2  & 0.14073   & -1.60(8)e-3  & 10048 \\ 
 1.1  & 0.142812 & -2.79(7)e-3  & 9978 \\ 
\hline
\end{tabular}
\caption{Parameter $\kappa$ used in the simulations, the PCAC mass at each $\beta_L=4/g_0^2$ and the number of measurements performed on the largest lattice.}
\label{table:kappac}
\end{table}

We note that in addition to the clover term, there are order $a$ improvement terms that can be added
to the action at the timelike boundaries of the lattice \cite{Karavirta:2010ef,Karavirta:2011mv}
and to the axial current correlator $f_A$ \cite{Luscher:1996vw}.
Since we have chosen to use the tree-level value for the clover coefficient $\csw$, improving the step scaling function only to the first order in $g^2$, we have consistently chosen to leave these improvements to the tree-level, where they have no effect. 

The simulations are done using the hybrid Monte Carlo (HMC) algorithm with the 2nd order Omelyan integrator \cite{Omelyan,Takaishi:2005tz}
and the chronological initial condition for the fermion matrix inversion \cite{Brower:1995vx}.
The length of the trajectory is fixed to 1 unit and the step size is tuned so that the acceptance rate is at least 80\%.
The measurements are taken after every trajectory and the numbers of trajectories generated using the largest lattice size are given in table \ref{table:kappac}.

We consider lattices of size $V=(aN)^4=L^4$ with Schr\"odinger functional boundary conditions. The spatial boundary conditions are periodic and at the $x_0=0$ and $x_0=L$ time slices the spatial gauge links are set to $U_k(x) = 1$.
The fermion fields are set to zero at the time boundaries
and have twisted periodic boundary conditions in the spacial directions:
\begin{align}
  \psi(x + L\hat k) = e^{i\pi/5}\psi(x).
\end{align}

The gradient flow coupling is defined by the action of a gauge field smoothed by a trivializing flow \cite{Luscher:2010iy,Luscher:2011bx,Fodor:2012td,Fritzsch:2013je}.
We define the field $B_{t,\mu}$ by the flow equation
\begin{align}
  \partial_t B_{t,\mu} &= -\frac{\delta S_G}{\delta B_{t,\mu}}, \\ 
  B_{0,\mu} &= A_\mu,
\end{align}
where $S_G$ is the action in equation \ref{plaqaction}, with the gauge field replaced by the flow field $B_{t,\mu}$. The coupling is then given by
\begin{align}
  &g^2 = \frac{t^2 \left < E(t) \right>}{N(t,a/L)},\\
  \ev{ E(t) }  &= \frac 14 \ev{ G_{\mu\nu}(t)G_{\mu\nu}(t) }.
\end{align}
We use the clover definition for the observable $G_{\mu\nu}$.

The coefficient $N(t,a/L)$ is chosen so that the gradient flow coupling coincides with the bare coupling at tree-level for each lattice size. Since fermion loops do not affect the coupling at tree-level, it can be conveniently measured from pure gauge simulations. We measure $N(t,a/L)$ using 35 000 configurations with $\beta_L=80$. The statistical error on the largest lattice is order per mille, and we include it in the analysis using the jackknife method described below.

We study the running of the coupling using finite size scaling. With the flow time fixed to $t=c^2L^2/8$ we vary lattice size $L$ and measure the response of the coupling.
The change is quantified using the step scaling function \cite{Luscher:1993gh}
\begin{align}\label{eq:sigma_def}
  &\Sigma(u,\A/L) = \left . g^2(g_0,sL/\A) \right|_{g^2(g_0,L/\A)=u}\\
  &\sigma(u) = \lim_{\A \rightarrow 0} \Sigma(u,\A/L)
\end{align}
In this study we choose $c=0.4$ and $s=2$.

\begin{table}
\centering
\begin{tabular}{llll}
\hline
 $\beta_L$ & $g^2$ & $\tau$    \\ \hline
\hline
 8     & 0.569(2) & 19.66  \\   
 4     & 1.194(5) & 6.115  \\ 
 2     & 2.42(1)   & 9.328  \\ 
 1.5  & 3.30(2)   & 10.48  \\ 
 1.3  & 3.93(3)   & 12.51  \\ 
 1.2  & 4.30(4)   & 14.02  \\ 
 1.1  & 4.73(5)   & 19.05  \\   
\hline
\end{tabular}
\caption{ The measured couplings and the autocorrelation times of the measurement and the number of measurements per jackknife block on the largest lattice size. }
\label{couplingandcorrelations}
\end{table}

Compared to the Sch\"odinger functional method, the gradient flow coupling tends to have good statistical accuracy allowing for shorter runs and larger lattice size. However, it tends to have long autocorrelation times. We measure the coupling after each HMC trajectory and find an autocorrelation time of $\sim20$. The values of the coupling and the autocorrelation times on the largest lattice are given in table \ref{couplingandcorrelations}. To propagate the error consistently we use the jackknife method with 40 blocks for each set of parameters, giving a block size significantly larger than the longest observed autocorrelation time.

We also note that, given the largest lattice sizes and couplings, topological freezing could be an issue. We measure the topological charge using the cooled gauge configurations at nonzero gradient flow time. After a period of thermalization, the topological charge settles to zero in all runs and never deviates for more than a couple of measurements. We include all generated configurations in the analysis.

\begin{figure} \center
  \includegraphics[width=0.4\textwidth,height=0.4\textwidth]{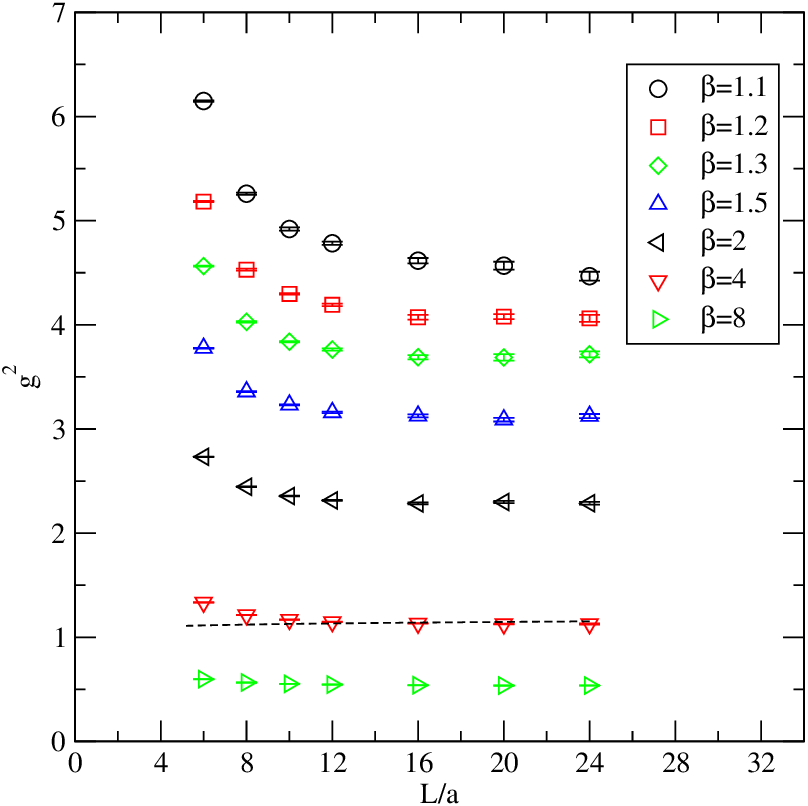}
\caption[b]{
  The measured values of the gradient flow coupling against $L/a$ at different values of $\beta=4/g_0^2$. The dashed line indicates the perturbative running coupling in continuum at $\beta=4$ at $L/a=24$.
}
\label{fig:coupling}
\end{figure}

The measured coupling is shown in figure \ref{fig:coupling} as a function of the lattice size. It is clear form the figure that the coupling deviates significantly from the expected slow running on the coarsest lattice. The gradient flow coupling, as defined here, produces significant discretization effects at order $a^2$. Several ways of alleviating them have been studied in \cite{Cheng:2014jba,Fodor:2014cpa} and \cite{Ramos:2015baa}. They also appear in the step scaling function at order $a^2$ and are removed by the continuum extrapolation. Higher order effects are small in comparison and cannot be distinguished from statistical errors at current accuracy.

\subsection{Interpolation in $g_0^2$}

\begin{figure*}
\includegraphics[width=0.4\textwidth,height=0.4\textwidth]{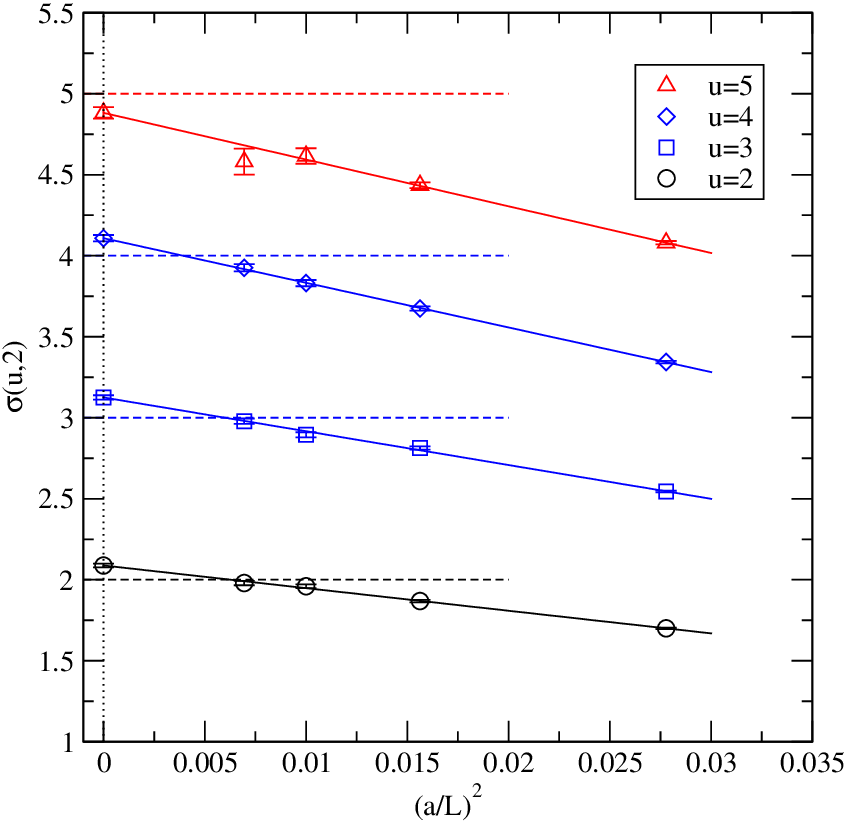}
\includegraphics[width=0.4\textwidth,height=0.4\textwidth]{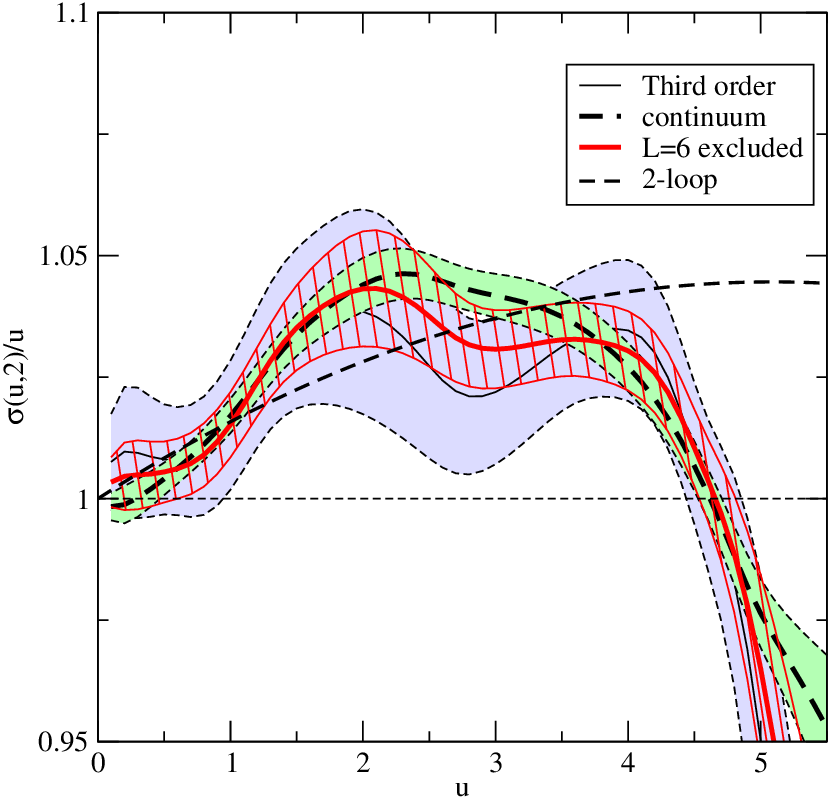}
\caption[b]{
 Left: The continuum extrapolation of the step scaling function at a few values of $u=g^2$. Right: The scaled continuum step scaling function $\sigma(u,2)/u$. The dashed line and the light green area depicts the second order continuum fit using all lattice sizes and the associated statistical error. The red curve with the red hashed area depicts the second order extrapolation with $L/a=6$ excluded and the solid black curve with the light blue area gives the third order fit.
}
\label{fig:sigmacont_int}
\end{figure*}

We use two different methods for taking the continuum limit. The first, the interpolation method, is more traditional and was first used in \cite{Bursa:2011ru}. It is based on parametrizing the data at each lattice size $L/a$ as a function of $g_0^2$ using an interpolating function. We fit the data to the polynomial function
\begin{align}\label{eq:sigma_int}
  &g^2(g_0^2) = g_0^2 \left ( 1 + \sum_{k=1}^{m} c_k g_0^{2k} \right )
\end{align}
The order of the polynomial, $m=6$, is found by minimizing the combined $\chi^2/d.o.f$ of the fit. This choice only leaves $1$ degree of freedom per lattice size. The high dimension of the fit function is largely due to the quick deviation of the results from the coarsest lattice from the tree level. We study the robustness of the fit by also running the analysis with $m=5$. The result is essentially unchanged and the difference is included in the reported systematic error. The combined $\chi^2/d.o.f$ of the fit is $\sim 1.5$.

The interpolating function allows us to calculate the step scaling function at any value of $u=g^2(g_0^2)$ and perform a continuum extrapolation using all the available lattice sizes.
We expect the lowest order discretization errors to be order $\A^2$ and fit the lattice step scaling function to
\begin{align}\label{eq:sigma_cont_int}
  &\Sigma(u,\A/L)  = \sigma(u) + c(u) \frac{\A^2}{L^2}.
\end{align}

We show the continuum extrapolation on the left in figure \ref{fig:sigmacont_int}. While the scaling is steep, the second order fit describes the data through most of the fit range. However, the $\chi^2/d.o.f$ of the fit exceeds $1$ at several points along the range, and has a maximum of $2.2$ at $u=4.6$. 
We have experimented with additional terms in the extrapolation, including term linear in $a$ fit, but these are not favored by the data.

The systematic error of the continuum fit is estimated by excluding the smallest lattice size, $L=6$ and by adding a third order term in the extrapolation. The continuum limit of the step scaling function is shown on the right in figure \ref{fig:sigmacont_int} together with the result with the smallest lattice size excluded and the third order extrapolation. The different fits agree up to $g^2\sim 5$. We find a fixed point at ${g^\ast}^2\approx 4.6$.

\subsection{Power Series Fit}

\begin{figure*}
\includegraphics[width=0.24\textwidth,height=0.24\textwidth]{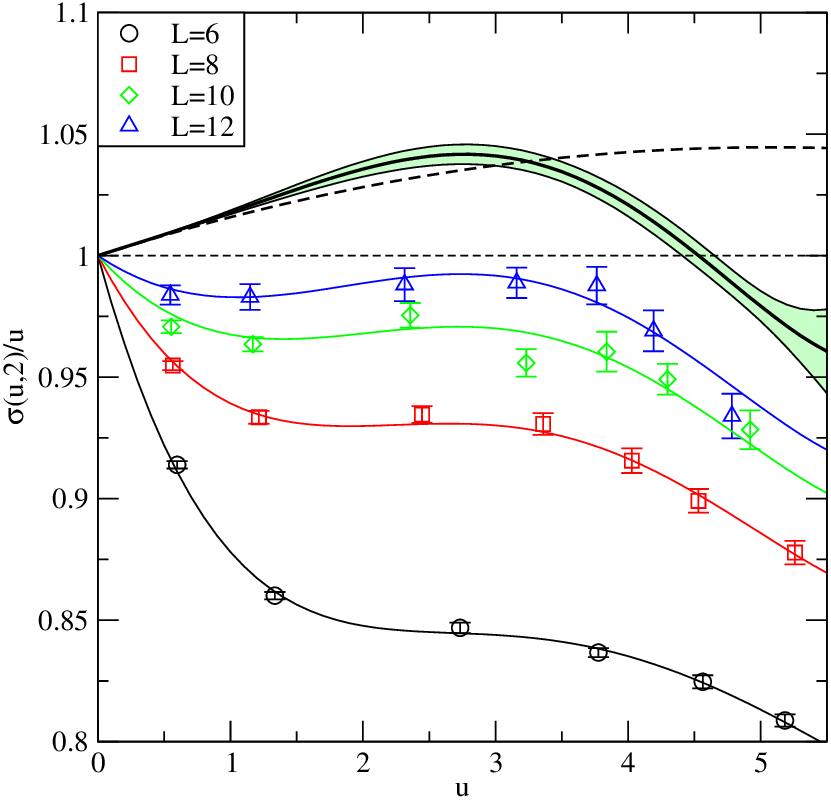}
\includegraphics[width=0.24\textwidth,height=0.24\textwidth]{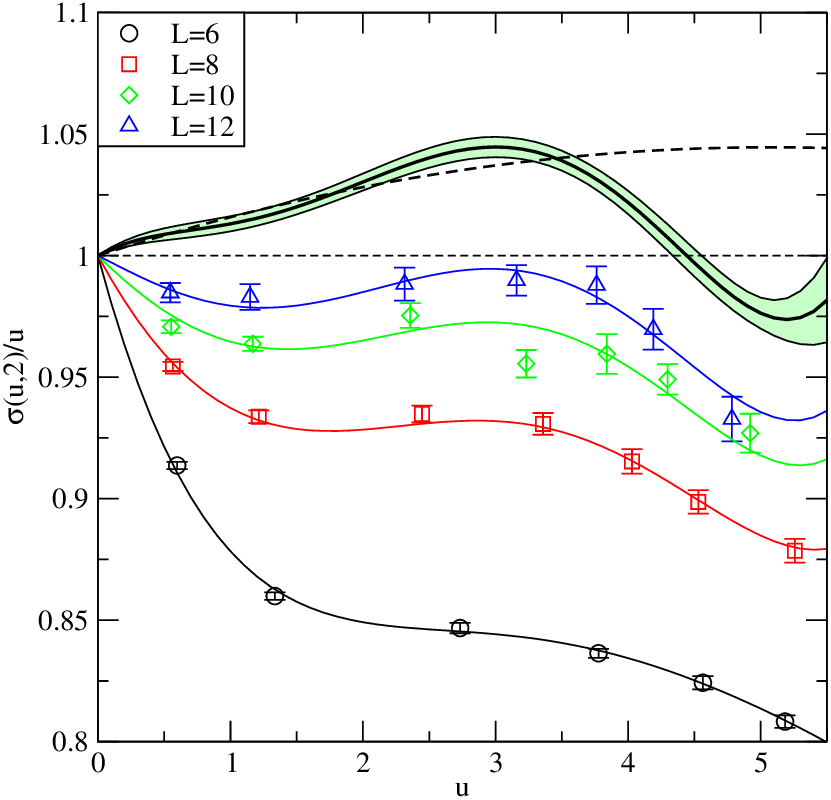}
\includegraphics[width=0.24\textwidth,height=0.24\textwidth]{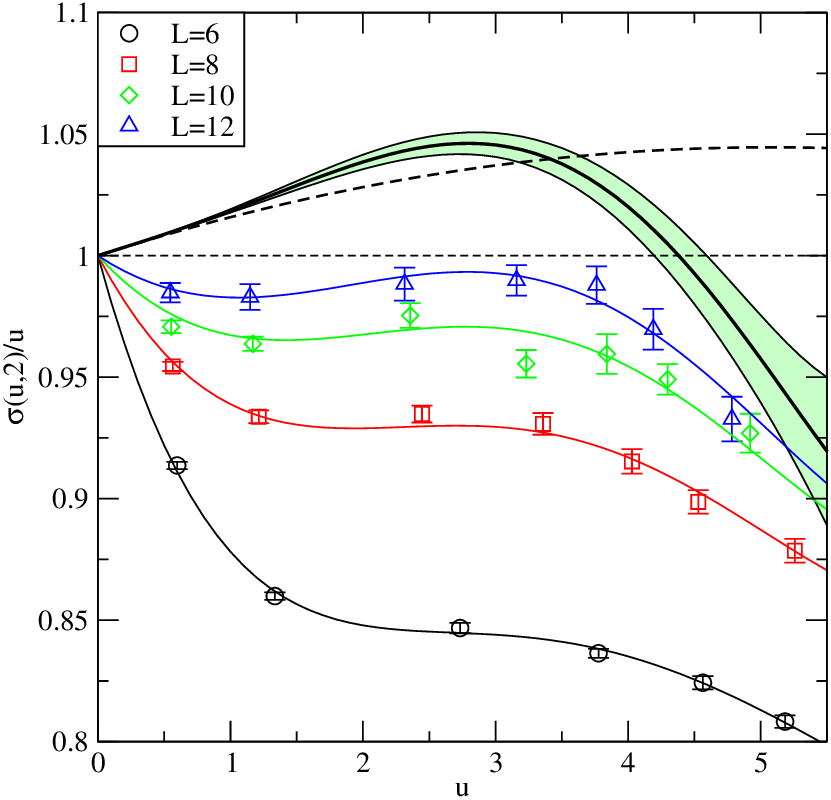}
\includegraphics[width=0.24\textwidth,height=0.24\textwidth]{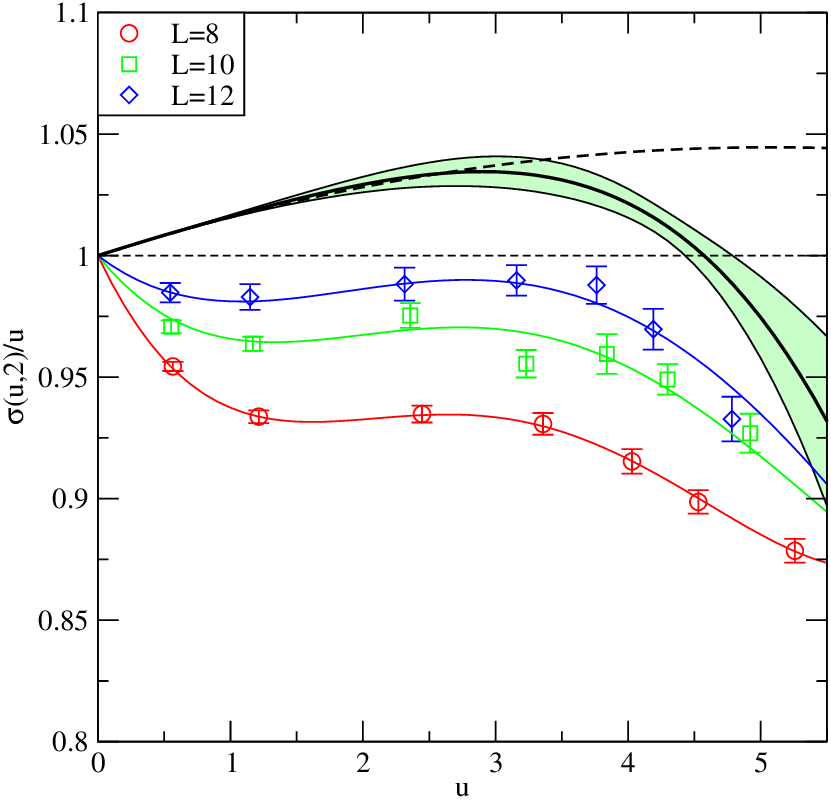}
\caption[b]{
 Power law fits to the step scaling function. From left to right: Constrained, $n_a=2$, $n=5$ and $n_f^2=5$; Unconstrained, $n_a=2$, $n=5$ and $n_f^2=5$; Constrained, $n=4$, $n_a=3$, $n_f^2=4$, and $n_f^3=2$; Constrained with smallest lattice excluded, $n=4$, $n_a=2$, $n_f^2=4$, and $n_f^3=2$.  The green band shows the statistical error of the continuum extrapolation.
}
\label{fig:sigmacont_pow}
\end{figure*}

The power series fit, also used in \cite{Rantaharju:2015yva}, allows us to leverage the data more efficiently.  We represent the step scaling function and the discretization errors as a power series in $u$ and $a$ and fit to all lattice data simultaneously. Since the step scaling function at each lattice size is a smooth function of the renormalized coupling, we can use a polynomial function to represent it with a relatively small number of parameters. The method also allows us to use the known universal two loop expansion of the $\beta$ function.

The fit function has the form
\begin{align} 
  &\frac{\sigma(u,2)}{u} =  1 + \sum_{i=1}^n c_i u^i \nonumber \\
  &\Sigma(u,2,a/L) = \sigma(u,2) + \sum_{k=2}^{n_a} f_k(u) \frac{a^k}{L^k} \label{eq.sigma_func_extrap}  \\
  &f_k(u) =  \sum_{l=1}^{n_f^k} c_{k,l} u^l. \nonumber
\end{align}
Here $c_i$ and $c_{k,l}$ are fit parameters. The fit can be further constrained by setting the parameters $c_1$ and $c_2$ to their known perturbative values.

In figure \ref{fig:sigmacont_pow} we show the continuum step scaling function and the fit to the lattice results. The parameters $n$ and $n_f^k$ are chosen to minimize the $\chi^2/d.o.f$ of the fits and the fit is stable against variations of these parameters. 
The first two panels in figure \ref{fig:sigmacont_pow}  show a constrained and an unconstrained fit with $n_a=2$. The $\chi^2/d.o.f$ are $\sim 1.2$ and $\sim 1.5$ respectively and the statistical errors are found using the jackknife method.
In both cases the running is slightly faster than perturbative below $g^2=3$. The running then becomes slower and there is an IRFP at $g^{*2}\approx 4.5(1)$.
The fit shown in the third panel in figure \ref{fig:sigmacont_pow} includes an $O(a^3)$ correction and the one in the fourth panel is done excluding the smallest lattice size. The  $\chi^2/d.o.f$ values are $\sim 1.5$ and $\sim 0.8$ and the fits show a fixed point at $g^{*2}=4.4(2)$ and $g^{*2}=4.6(2)$ respectively. 

As our final result for the location of the IRFP we quote the value obtained from the constrained power law fit with $n_a=2$, $n=5$ and $n_f^2=5$, 
${g^*}^2 = 4.5(1)^{+0.3}_{-0.3}$. The first error is statistical and the second includes the range of results from different estimates of the continuum limit and from varying the parameters at each step. At the infrared fixed point, we find the anomalous dimension of the coupling, $$\gamma_g^* = \lim_{g\rightarrow {g^*}} \frac{\beta(g)}{g} = 0.3(0.05)^{+0.15}_{-0.08}. $$

\section{The Mass Anomalous Dimension} \label{gamma}

In order to study the mass anomalous dimension in the gradient flow scheme we use the measurements first published in \cite{Rantaharju:2015yva}. We summarize the method here and refer the interested reader to \cite{Rantaharju:2015yva} and to \cite{Capitani:1998mq, DellaMorte:2005kg}, where the method is described in more detail. 

The mass anomalous dimension is measured from the running of the pseudoscalar density renormalization constant. It can be measured from the boundary to bulk pseudoscalar correlator $f_P(x_0)$ normalized by the boundary to boundary correlator $f_1$,
\begin{align}
Z_P(L) = \frac{\sqrt{3 f_1} }{f_P(L/2)}
\label{Zp}
\end{align}
The pseudoscalar step scaling function is then defined by
\begin{align}
 \Sigma_P(u,2,L/a) &= \left . \frac{Z_P(g_0^2,2L/a)}{Z_P(g_0^2,L/a)}\right|_{g^2(g_0,L/\A)=u}\\
 \sigma_P(u,2) &= \lim_{a\rightarrow 0} \Sigma_P(u,2,L/a)
\end{align}
At the fixed point the pseudoscalar step scaling function is related to the mass anomalous dimension as
\begin{align} \label{estimate}
  \gamma^* = \bar\gamma({g^*}^2),\,\,\, \bar \gamma(u) = -\frac{\log{\sigma_P(u,2)}}{\log(2)}
\end{align}

\begin{figure*}
 \includegraphics[width=0.4\textwidth,height=0.4\textwidth]{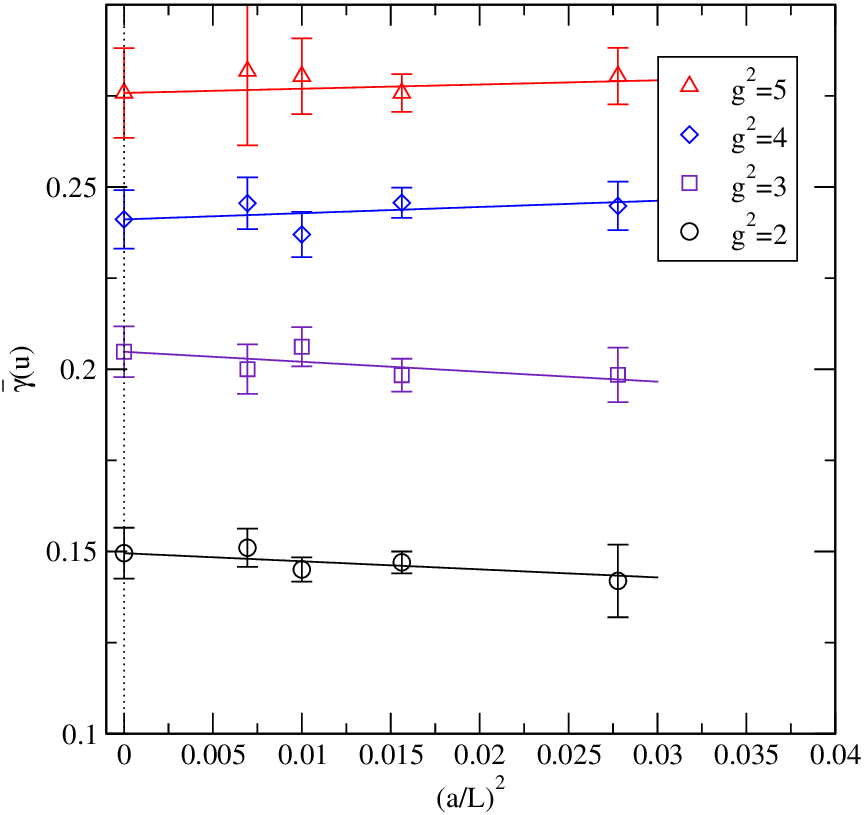}
 \includegraphics[width=0.4\textwidth,height=0.4\textwidth]{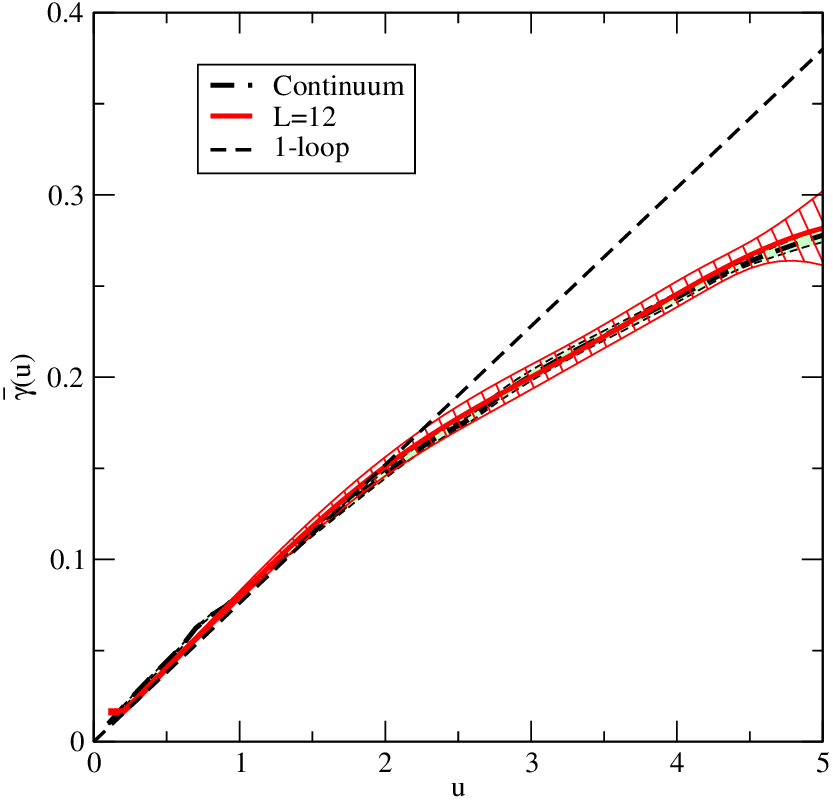}
\caption[b]{
 Left: The continuum extrapolation of the estimator for the anomalous dimension $\bar\gamma(u)$ at a few values of $u$. Right: The continuum limit of $\bar\gamma(u)$
}
\label{fig:gamma_cont_int}
\end{figure*}

We use the interpolation method to take the continuum limit. First, we fit the pseudoscalar step scaling function to the interpolating function
\begin{align}\label{eq:zp_int}
  &Z_p(g_0^2) = 1 + \sum_{k=1}^{m} c_k g_0^{2k}
\end{align}
The $\chi^2/d.o.f$ of the fit is minimized by $m=5$ with the value $\approx 0.930$. The continuum limit is then found for each value of $u$ by fitting to
\begin{align}
 \gamma_L(u,2,L/a) &= \bar\gamma(u,2) + c(u) \frac{a^2}{L^2}, \\
 \gamma_L(u,2,L/a)  &=\left .I -\frac{\log{\Sigma_P(g_0^2,2,L/a)}}{\log(2)} \right|_{g^2(g_0,L/\A)=u} 
\end{align}

The continuum limit of the scaled quantity $\bar\gamma(u)$ is shown in figure \ref{fig:gamma_cont_int} along with the result form the largest lattice size. The continuum extrapolation is milder than for the step scaling function and the result at $L/a=12$ agrees well with the continuum limit. The result follows the two loop perturbative result up to $u=2$ and has a smaller value after that. At the fixed point we find the value $\gamma^* = 0.263(4)^{+0.012}_{-0.015}$. The first error estimate is purely statistical and the second includes an estimate of systematic errors and the uncertainty of the location of the IRFP.

\section{Conclusions} \label{conclusions}

We have presented a lattice study of the SU(2) gauge theory with 2 flavors of fermions in the adjoint representation of the gauge group. We have measured both the running coupling and the mass anomalous dimension in the gradient flow scheme using the same lattice formulation of the theory as the one used in \cite{Rantaharju:2015yva} to study the model in the Schr\"odinger functional scheme. The discretization effects present in the lattice model should therefore be the same in both studies and any difference should result from discretization effects and systematic errors in the measurables.

We have measured the coupling at a larger lattice size than before and as a result have a better control over the continuum limit. The definition of the gradient flow coupling used introduces a large discretization effect of order $a^2$. We observe steep approach to the continuum, but find that the $O(a^2)$ extrapolation describes the data well and expect higher order effects to be small in comparison. Our results confirm the existence of an infrared fixed point and we find ${g^*}^2 = 4.6^{+0.2}_{-0.3}$. A fixed point was also found in \cite{Hietanen:2009az,Bursa:2009we,DeGrand:2011qd} and \cite{Rantaharju:2015yva} in the Schr\"odinger functional scheme.

The mass anomalous dimension is a scheme independent quantity and can be compared between different studies.
We find the value $\gamma^\ast = 0.263^{+0.012}_{-0.015}$ at the fixed point. The value found in \cite{Rantaharju:2015yva} is smaller, $\gamma^\ast\simeq 0.2 \pm 0.03$. The behavior of the estimator $\bar\gamma(u)$ is similar in both studies and the difference likely arises from the uncertainty of the location of the IRFP. Our result is compatible with the value obtained in \cite{DeGrand:2011qd}, $\gamma^\ast=0.31(6)$. In \cite{Patella:2012da} $\gamma^*=0.37(2)$ was obtained using a different method.

\acknowledgments 
We thank K. Rummukainen and A. Ramos for useful discussion.
This work is supported by the Danish National Research Foundation grant number DNRF:90.
The simulations were performed on the k-computer at Riken AICS in Kobe, Japan.

\end{document}